\documentclass[conference]{IEEEtran}
\IEEEoverridecommandlockouts

\def\BibTeX{{\rm B\kern-.05em{\sc i\kern-.025em b}\kern-.08em
    T\kern-.1667em\lower.7ex\hbox{E}\kern-.125emX}}

\usepackage{amsmath}
\usepackage{amsfonts}
\usepackage{algorithm}
\usepackage{algorithmic}
\usepackage{graphicx}
\usepackage{textcomp}
\usepackage{listings}
\usepackage[normalem]{ulem}
\usepackage{booktabs}
\usepackage{multirow}
\usepackage{subcaption}
\usepackage[dvipsnames]{xcolor}

\usepackage{cite}
\usepackage{amssymb}

\lstdefinestyle{mystyle}{
    commentstyle=\color{codegreen},
    keywordstyle=\color{magenta},
    stringstyle=\color{codepurple},
    basicstyle=\ttfamily\footnotesize,
    breakatwhitespace=false,         
    breaklines=true,                 
    captionpos=b,                    
    keepspaces=true,                 
    numbers=left,                    
    numbersep=5pt,                  
    showspaces=false,                
    showstringspaces=false,
    showtabs=false,                  
    tabsize=2,
    xleftmargin=2em,
    frame=single
}

\renewcommand{\r}[1]{{{{\color{red}#1}}}} 
\renewcommand{\b}[1]{{{{\color{black}#1}}}}
\newcommand{\g}[1]{{{{\color{Green}#1}}}}
\newcommand{\y}[1]{{{{\color{Peach}#1}}}}
\newcommand{\toolname}{RampoNN~}

\newcommand{\nn}{\mathcal{N\!N}}

\newtheorem{theorem}{\textbf{Theorem}}[section]
\newtheorem{problem}[theorem]{\textbf{Problem}}
\newtheorem{definition}[theorem]{\textbf{Definition}}
\renewcommand{\baselinestretch}{0.94}

\begin{document}

\title{\toolname\!: A Reachability-Guided System Falsification for Efficient Cyber-Kinetic Vulnerability Detection}

\author{\IEEEauthorblockN{Kohei Tsujio, Mohammad Abdullah Al Faruque, and Yasser Shoukry}
\IEEEauthorblockA{\textit{Department of Electrical Engineering and Computer Science} \\
\textit{University of California, Irvine}\\
\{ktsujio, alfaruqu, yshoukry\}@uci.edu}
}
\maketitle
\begin{abstract}

Detecting kinetic vulnerabilities in Cyber-Physical Systems (CPS)---vulnerabilities in control code that can precipitate hazardous physical consequences—is a critical challenge. This task is complicated by the need to analyze the intricate coupling between complex software behavior and the system's physical dynamics. Furthermore, the periodic execution of control code in CPS applications creates a combinatorial explosion of execution paths that must be analyzed over time, far exceeding the scope of traditional single-run code analysis.

This paper introduces \toolname\!\!, a novel framework that systematically identifies kinetic vulnerabilities given the control code, a physical system model, and a Signal Temporal Logic (STL) specification of safe behavior. \toolname first analyzes the control code to map the control signals that can be generated under various execution branches. It then employs a neural network to abstract the physical system's behavior. To overcome the poor scaling and loose over-approximations of standard neural network reachability, \toolname uniquely utilizes Deep Bernstein neural networks, which are equipped with customized reachability algorithms that yield orders of magnitude tighter bounds. This high-precision reachability analysis allows \toolname to rapidly prune large sets of guaranteed-safe behaviors and rank the remaining traces by their potential to violate the specification. The results of this analysis are then used to effectively guide a falsification engine, focusing its search on the most promising system behaviors to find actual vulnerabilities.

We evaluated our approach on two case studies: a PLC-controlled water tank system and a switched PID controller for an automotive engine. 
The results demonstrate that \toolname leads to acceleration of the process of finding kinetic vulnerabilities by up to $98.27\%$ and superior scalability compared to other state-of-the-art neural network reachability and falsification methods.

\end{abstract}

\section{INTRODUCTION}

\label{sec:introduction}
Cyber-Physical Systems (CPS) form the backbone of modern critical infrastructure, including autonomous vehicles, industrial control systems, and medical devices. Ensuring the safety and security of these systems is paramount, as a software failure or vulnerability can lead to catastrophic physical consequences. Verifying the complex control programs within CPS, however, presents a formidable challenge. The core difficulty stems from the tight coupling between their discrete software logic and the continuous physical dynamics they govern. Unlike purely software systems, the behavior of a CPS is not solely determined by its control logic; it is a complex, continuous interplay where software commands influence physical processes, and the resulting physical state, in turn, provides feedback that dictates subsequent software execution. This fundamental feedback loop renders traditional software verification techniques, which are ``blind'' to physical dynamics, inadequate for reasoning about system safety.

\begin{figure}
    \centering
    \includegraphics[width=\linewidth]{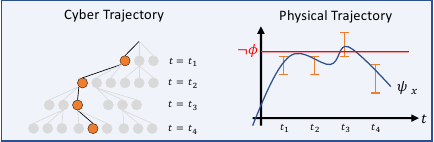}
    \vspace{-6mm}
    \caption{
    An example of cyber-kinetic vulnerability that shows both the physical trajectory (right) violating some safety constraint $\varphi$ and the corresponding cyber trajectory (left).
    The cyber trajectory tree illustrates the combinatorial explosion where 
    each branch represents a different control decision.
    \vspace{-4mm}
    }
    \label{fig:intro}
\end{figure}

To exacerbate the problem, this cyber-physical coupling introduces a critical temporal dimension. The response of a physical system to a software command is not instantaneous; its dynamics evolve and accumulate over time. Consequently, a single, flawed execution of the control loop might not, in itself, reveal a safety violation. Instead, the hazardous effects of a sequence of seemingly benign control decisions can accumulate, manifesting as a catastrophic failure only after a significant delay. Therefore, simply analyzing a single execution path of the software—as is common in traditional verification—is fundamentally insufficient. A comprehensive analysis must consider the cumulative effects of software execution across multiple time horizons, necessitating the exploration of all possible sequences of software execution paths.

However, as illustrated in Figure~\ref{fig:intro}, this requirement for long-horizon analysis immediately collides with the challenge of combinatorial explosion. The behavior of a CPS control program can be modeled as a ``cyber trajectory'' tree, where each path from the root represents a possible sequence of control decisions over time. As the time horizon ($H$) increases, the number of possible cyber trajectories grows exponentially (the $k^H$ problem, where $k$ is the number of branches at each decision point). This exponential growth makes exhaustive verification computationally infeasible for any non-trivial system.

\subsection{Related Work}

\noindent\textbf{$\bullet$ Fuzzing for Cyber–Physical Systems:}
Fuzzing has been applied to CPS to automatically explore input spaces and uncover unsafe behaviors.  
Frameworks such as ~\cite{9797531,10.1145/3395363.3397376,pgfuzz} use system-level feedback (e.g., coverage, policy guidance, or physical signal metrics) to guide the mutation of input trajectories.  
Similar CPS-oriented fuzzing approaches, such as ~\cite{10.1145/3548606.3560558}, use domain-specific objectives derived from physical outcomes to reveal safety-critical corner cases.  
These methods demonstrate the value of adaptive input generation but largely treat the controller and the physical dynamics as a black box, without leveraging structural information about the control software or formal overapproximations of the system’s continuous behavior.  
As a result, they may repeatedly explore input regions that are infeasible for the underlying software or physically unreachable under the system dynamics.

\noindent\textbf{$\bullet$ Hybrid-System Verification and Reachability:}
Reachability analysis provides formal guarantees for continuous or hybrid systems by computing overapproximations of all possible system trajectories under bounded inputs.  
Tools such as ~\cite{10.1007/978-3-642-14295-6_17,10.5555/2032305.2032335,flow*} have demonstrated the ability to analyze nonlinear dynamics and verify safety properties in a mathematically rigorous manner.  
Advanced methods based on zonotopes~\cite{10.1007/978-3-540-31954-2_19}, interval analysis~\cite{10.1007/978-3-030-53288-8_27}, or high-performance reachability engines such as ~\cite{10.1145/3049797.3049808,10.1145/3302504.3311804} further extend scalability.  
However, these approaches typically assume a fixed hybrid model with predetermined discrete transitions and do not incorporate the behavior of complex control programs or data-dependent branching.  
Moreover, while reachability can provide provable overapproximations of the continuous dynamics, it does not directly address whether the required control actions can be generated by real software, leading to conservativeness when used in isolation.

\noindent\textbf{$\bullet$ Verification of Neural Networks and Learning-Enabled Systems:}
A growing body of work develops formal verification techniques for neural networks, including SMT-based solvers such as ~\cite{10.1007/978-3-319-63387-9_5,10.1007/978-3-030-25540-4_26}, abstract-interpretation methods~\cite{8418593}, and output range estimation techniques~\cite{10.1007/978-3-319-77935-5_9,8318388}.  
Integration with CPS verification has also been explored: ~\cite{verisig,verisig2,nnv,10.1007/978-3-031-37703-7_19,reachnn,reachnn_star} combine neural network reachability with continuous dynamics to analyze safety of systems with neural controllers.  
Other work develops principled approaches to analyze neural network–based feedback loops~\cite{10.1145/3302504.3311807}, compose CPS reasoning with machine-learning components~\cite{10.1007/s10817-018-09509-5}, or embed formal specifications directly into neural architectures~\cite{stl2nn}.  
These methods provide important foundations for reasoning about learned components but do not consider general control software with complex branching, nor do they focus on guiding falsification by combining program-level constraints with reachable set information.

\noindent\textbf{$\bullet$ Falsification and Search-Based Testing for CPS:}
Search-based falsification uses quantitative semantics of Signal Temporal Logic~\cite{stl} (STL) to guide the search toward trajectories that violate system specifications.  
Tools such as ~\cite{s-taliro} and efficient STL monitoring algorithms~\cite{10.5555/2958031.2958095} enable robustness-based optimization over input trajectories.  
Recent frameworks such as ~\cite{10.1007/978-3-030-25540-4_25} extend falsification to systems with machine-learning components through structured scenario sampling.  
While these methods provide powerful search-based strategies, they do not combine formal overapproximations of the continuous dynamics with software-level structural information, leaving a gap in ruling out unreachable or unrealizable behaviors.

Despite the progress in aforementioned areas, existing methods typically focus on only one aspect of the system at a time.  
Consequently, they may explore input behaviors that are physically unreachable, analyze continuous evolutions that the software cannot realize, or rely on repeated simulations and unguided falsification without structural mechanisms to eliminate infeasible regions of the search space.  
A scalable analysis therefore requires combining the strengths of these approaches.

\subsection{Our Contribution}

To overcome the fundamental limitation of combinatorial explosion in the cyber-trajectory search space, we introduce \toolname\!\!, a novel framework that significantly accelerates vulnerability detection in CPS. Instead of relying solely on computationally expensive ``blind'' falsification, \toolname introduces a verifiable, ``reachability-based guided'' falsification strategy.

Our core innovation is the insight that this pruning can be made both tight and computationally efficient by composing neural network models of both the system dynamics (\textit{DynamicsNN}) and the STL specification (\textit{STL2NN}) using Deep Bernstein (DeepBern) networks. This specific neural architecture, unlike standard ReLU-based networks, is amenable to high-precision reachability analysis. It allows us to compute tight, mathematically rigorous bounds on the system's robustness score for entire sets of abstract cyber trajectories at once.

\toolname uses these bounds to systematically explore the abstract trajectory tree, soundly pruning provably safe subtrees and drastically reducing the analysis scope. The computationally expensive falsification engine is then guided to focus only on the remaining, dramatically reduced set of uncertain or unsafe trajectories.

This paper makes the following specific contributions:
\begin{enumerate}
\item \textbf{DeepBern-Net Based Robustness Analysis:} We propose the integration of \textit{DynamicsNN} (a neural network that models the physical dynamics) with an analytically constructed neural network \textit{STL2NN} (modeling the STL specification). Our key technical contribution is the use of DeepBern-Nets for both components. This architecture avoids the piecewise-linear branching inherent in ReLU-based models, which are intractable for deep reachability, and instead enables the efficient computation of tight, global bounds on the system's end-to-end robustness.

\item \textbf{Reachability-Guided Falsification:} We introduce the \toolname~framework, which integrates this verifiable pipeline into a tree-based exploration of abstract cyber trajectories. The framework uses the computed robustness bounds to classify abstract trajectories as SAFE, UNSAFE, or UNCERTAIN, and soundly prunes the SAFE subtrees. This approach transforms the intractable $k^H$ search problem into a manageable, guided search, focusing expensive falsification resources only where they are most needed.

\item \textbf{Demonstrated Scalability and Enhanced Vulnerability Detection:} We provide an empirical evaluation on two complex CPS benchmarks: a PLC Water Tank model and an Automotive Engine with a Switched PID controller. We demonstrate that \toolname successfully uncovers complex, deeply-nested cyber-kinetic vulnerabilities that are missed by existing falsification methods and other NN-reachability approaches due to their inherent scalability limitations.

\end{enumerate}
\section{PROBLEM DEFINITION}

\subsection{Notation and Preliminaries}
\noindent \textbf{$\bullet$ Notation.}
We use the symbols $\mathbb{N}$, $\mathbb{R}$, and $\mathbb{B}$ to denote the set of natural, real, and Boolean numbers, respectively. We use $\land$, $\lor$, and $\lnot$ to represent the logical AND, OR, and NOT operators, respectively.

\noindent \textbf{$\bullet$ Bernstein polynomials.}
Bernstein polynomials form a basis for the space of polynomials on a closed interval. These polynomials have been widely used in various fields
due to their unique properties and intuitive representation of functions.

A general polynomial of degree $n$ in Bernstein form on the interval $[l, u]$ can be represented as:

\begin{align}
\label{eq:bern_form}
P_n^{[l,u]}(x) = \sum_{k=0}^n c_k b_{n,k}^{[l,u]}(x), \qquad x \in [l,u]
\end{align}
where $c_k \in \mathbb{R}$ are the coefficients associated with the Bernstein basis $b_{n,k}^{[l,u]}(x)$, defined as:
\begin{align*}
\label{eq:bern_basis}
b_{n,k}^{[l,u]}(x) = \frac{\binom{n}{k}}{(u-l)^n}(x-l)^k (u - x)^{n - k},
\end{align*}
with $\binom{n}{k}$ denoting the binomial coefficient. The Bernstein coefficients $c_k$ determine the shape and properties of the polynomial $P_n^{[l,u]}(x)$ on the interval $[l, u]$. It is important to note that unlike polynomials represented in power basis form, the representation of a polynomial in Bernstein form depends on the domain of interest $[l, u]$ as shown in equation~\eqref{eq:bern_form}.

\noindent \textbf{$\bullet$ DeepBern-Nets: Neural Networks with Bernstein activation functions.}
In this paper, we make use of a special neural network architecture called DeepBern-Nets~\cite{DeepBern}, feed-forward NNs with Bernstein polynomials as non-linear activation functions $\sigma$. Like standard feed-forward NNs, DeepBern-Nets consist of multiple layers, each consisting of linear weights followed by non-linear activation functions. Unlike conventional activation functions (e.g., ReLU, sigmoid, tanh, ..), Bernstein-based activation functions are parametrized with learnable Bernstein coefficients $c = c_0, \ldots, c_n$, i.e.,
\begin{align*}
\sigma(x;l,u,c) = \sum_{k=0}^n c_k b_{n,k}^{[l,u]}(x), \qquad x \in [l,u] \text{,}
\end{align*}
where $x$ is the input to the neuron activation, and the polynomial degree $n$ is an additional hyper-parameter of the Bernstein activation and can be chosen differently for each neuron. 

\noindent \textbf{$\bullet$ Forward Reachability of DeepBern-Nets}
Let $\mathcal{X} \subseteq \mathbb{R}^{n}$ be a set of possible input vectors to the neural network $\mathcal{N\!N}$. We define the \emph{forward reachable set} of $\mathcal{N\!N}$ with respect to $\mathcal{X}$ as
\[
\mathcal{R}_{\mathcal{N\!N}}(\mathcal{X}) \;=\; \{ \mathcal{N\!N}(x) \,\mid\, x \in \mathcal{X} \} \;\subseteq\; \mathbb{R}^{m}.
\]
Since exact reachable set computation is a typically heavily time-consuming process, existing algorithms focus on producing a conservative yet tight bounding set $\widehat{\mathcal{R}}$ such that:
\[
\mathcal{R}_{\mathcal{N\!N}}(\mathcal{X}) \;\subseteq\; \widehat{\mathcal{R}} \;\subseteq\; \mathbb{R}^m,
\]
where $\widehat{\mathcal{R}}$ can be represented in a tractable form (e.g., hyperrectangles or Bernstein polynomial bounding regions) that enables efficient intersection or emptiness checks.

Using Bernstein activation functions in neural network architecture allows us to compute these bounds more efficiently and precisely than traditional activation functions' networks even for neural networks with millions of parameters~\cite{DeepBern}. This is particularly important for our purpose, as tighter bounds lead to more accurate classification of cyber-kinetic vulnerabilities.

\subsection{CPS and Control Program}
We consider a discrete-time CPS defined by:
\begin{align*}
    x_{t+1} &= \nn_f(x_t, u_t) + \epsilon_t, \quad
    u_t = g(x_t),
\label{eq:NNf}    
\end{align*}
where $x_t \in \mathbb{R}^n$ is the state of the system at time $t \in \mathbb{N}$ and $u_t \in \mathbb{R}^m$ is the action at time $t$.
The dynamics of the system is assumed to be captured by a neural network  $\mathcal{N\!N}_f: \mathbb{R}^{n} \times \mathbb{R}^{m} \to \mathbb{R}^{n}$ while the term $\epsilon_t$ captures the error between the trained neural network $\nn_f$ model and the actual system. An upper bound $\bar{\epsilon}$ on this error $\epsilon$ is assumed to be given.
We assume that the system is controlled by a control program (software) $g:\mathbb{R}^n \rightarrow \mathbb{R}^m$. Without loss of generality, any software program can be decomposed using so-called path constraints~\cite{de2021symbolic}:
\begin{align*}
    g(x) = & [c^{(1)}(x) \Rightarrow u = g^{(1)}(x)]
    && \land  [c^{(2)}(x) \Rightarrow u = g^{(2)}(x)] \\
    & \qquad \hdots\hdots
    && \land [c^{(k)}(x) \Rightarrow u = g^{(k)}(x)],
\end{align*}
where $c^{(i)}:\mathbb{R}^n \rightarrow \mathbb{B}$ is the $i$th path constraint of the program $g$, and $g^{(i)}:\mathbb{R}^n \rightarrow \mathbb{R}^m$ is the path function, i.e., the function that is executed at the $i$th path of $g$.
\b{In this paper, we assume the controller is a state-feedback program, so that both the path decision and the control action at time $t$ depend only on the current state $x_t$. Consequently, the path-specific control bounds obtained by static analysis are also state-based. Finite-memory controllers can be handled by augmenting the state with controller memory variables, at the cost of a larger state dimension and increased analysis complexity.}
\begin{definition}[Cyber Trajectories]
A trajectory of the control program's internal states and decisions is referred to as a \textbf{cyber trajectory}.
We define a cyber trajectory as the sequence of \emph{path} that the control program encounters over a finite time horizon $H$.
Given a control program $g$  with  $k$  paths, a sequence:
\[\psi_p = \{p_t\}_{t=0}^H,~ p_t \in \{1,\dots,k\},\] is called a cyber trajectory of the physical system $\nn_f$ if there exists a corresponding trajectory of the physical system:
\[ \psi_x = \{x_t\}_{t=0}^H ~ \text{with} ~ x_{t+1} = \nn_f\bigl(x_t,\,g(x_t))\bigr) + \epsilon_t \]
s.t.
\[ p_t = \texttt{PATH}\bigl(x_t\bigr)  ~\text{for all}  ~t \in \{0,\dots,H\},\]
\[ \texttt{PATH}(x) = i \Longleftrightarrow c^{(i)}(x) = \texttt{True},\]
where $c^{(i)}$ is the $i$th path constraint of the program.
\end{definition}

\subsection{Cyber-Kinetic Vulnerabilities}
We are interested in enumerating all cyber trajectories that can lead to physical system trajectories that violate a safety requirement $\varphi$. In this paper, we assume that the safety requirement $\varphi$ is captured by an STL formula.
For the formal definition of STL syntax and semantics, we refer the reader to~\cite{maler2004monitoring}. Using this notation, the problem of interest can be defined as follows. 
\begin{problem}[Enumeration of Cyber-Kinetic Vulnerabilities]
Given an STL formula $\varphi$, a controller program $g$ with $k$ paths, and a physical system model $\nn_f$. Find the set $\Psi_p^\varphi$ of cyber trajectories that may lead to cyber-kinetic vulnerabilities, i.e., $\Psi_p^\varphi$ is defined as:
\begin{align*}
\Psi_p^\varphi = &\Big\{ \{p_t\}_{t = 0}^H \in \{1,\ldots,k\}^{H+1} \; \vert \;  \nonumber \\
& \exists \psi_x = \{x_t\}_{t = 0}^H . \big[ 
x_{t+1} = \nn_f(x_t, g(x_t)) + \epsilon_t \ (t < H), \nonumber\\ 
& \psi_x \not \models \varphi, \; p_t = \texttt{PATH}(x_t), \forall t \in \{0, \ldots, H\} \big]
\Big\}.
\end{align*}
\label{prob:cyber_kinetic}
\end{problem}
In other words, $\psi_p$ is considered a cyber-kinetic vulnerability if there
exists an associated trajectory of the physical system $\psi_x$ that violates
the safety requirement $\varphi$.

\begin{figure}[t]
    \centering
    \includegraphics[width=1\linewidth]{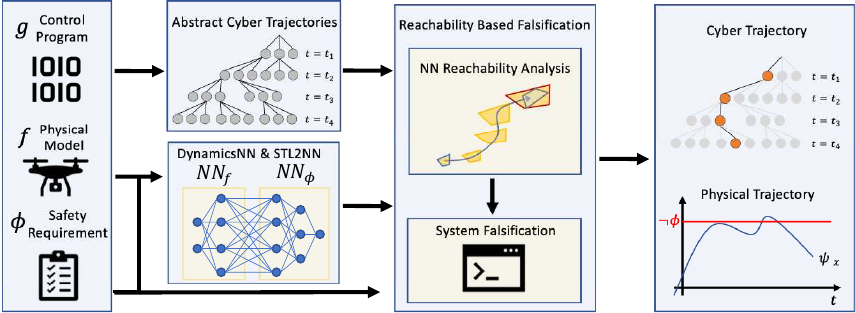}
    \caption{\toolname Framework Overview. \vspace{-3mm}}
    \label{fig:framework_overview}
\end{figure}

\section{FRAMEWORK}
\label{sec:framework}

\subsection{Overview of the \toolname Framework}
The primary obstacle in cyber-kinetic vulnerability detection is the exponential growth in the number of execution paths requiring analysis. For a control program with $k$ distinct paths and a time horizon of $H$, there exist $k^H$ possible cyber trajectories. Exhaustively analyzing each trajectory, for example via simulation or symbolic execution, quickly becomes computationally intractable as $H$ increases.

To tackle this combinatorial explosion, our approach is founded on two key ideas:
\begin{itemize}
    \item \textbf{Abstraction of Cyber Trajectories:} Instead of analyzing individual paths, we introduce a method to group concrete cyber trajectories into a manageable number of abstract cyber trajectories. Each abstraction represents a set of possible control flows.

    \item \textbf{Search Space Pruning using DeepBern-Nets:} Second, we leverage the cyber trajectory abstraction to perform neural network reachability analysis on $\nn_f$ and $\varphi$. This analysis computes provable bounds on the system's behavior for an entire abstract set of physical trajectories at once. Thanks to the DeepBern-Nets, we can compute tight reachable sets efficiently. Based on these bounds and the quantitative semantics of the Signal Temporal Logic (STL) specification $\varphi$, we classify each abstract set as: 1) \g{SAFE} (guaranteed to satisfy $\varphi$), 2) \r{UNSAFE} (guaranteed to violate $\varphi$), or 3) \y{UNCERTAIN}.
\end{itemize}

This classification achieves our primary goal: we can prune the \g{SAFE} sets entirely, while using the classification of the \r{UNSAFE} and \y{UNCERTAIN} sets to create a targeted, ``guided'' falsification strategy. This allows us to direct the computationally expensive falsification engine to focus its resources only on the most promising regions of the search space, rather than searching blindly.

The \toolname framework, illustrated in Figure~\ref{fig:framework_overview}, implements this strategy through four main components:

\begin{enumerate}
\item \textbf{Abstract Cyber Trajectory Generation}: This component analyzes the control code to generate a tree-based abstraction. Each node or path in this tree represents a set of concrete execution paths, effectively partitioning the $k^H$ search space into a smaller, finite set of abstract cyber trajectories.

\item \textbf{\textit{DynamicsNN} and \textit{STL2NN}}: We use two specialized neural networks. The \textit{DynamicsNN} $\nn_f$ is an input to our framework and is assumed to represent a system model capable of predicting the evolution of the physical system's state based on the current state and control actions. \textit{STL2NN} $\nn_\varphi$ is a neural network that will be automatically constructed by \toolname to compute the quantitative robustness score of a given state sequence with respect to the STL specification $\varphi$. Both $\nn_f$ and $\nn_\varphi$ are DeepBern-Nets.

\item \textbf{NN Reachability Analysis}: This component performs forward reachability analysis over the cascaded \textit{DynamicsNN} and \textit{STL2NN} models. This analysis computes provable output bounds on the final robustness score for each abstract cyber trajectory, classifying it as \g{SAFE}, \r{UNSAFE}, or \y{UNCERTAIN}.

\item \textbf{Guided System Falsification}: This component uses the results from the reachability analysis to prioritize the search for a concrete counterexample. The falsifier focuses on the \y{UNCERTAIN} and \r{UNSAFE} abstract trajectories to efficiently discover a specific cyber-kinetic vulnerability in the original, non-abstracted system.
\end{enumerate}

In the following subsections, we describe each component in detail and explain how they interact within our framework to efficiently detect cyber-kinetic vulnerabilities.

\subsection{Abstract Cyber Trajectory Generation}
\label{sec:two_level_abstraction}

Similar to~\cite{kohei}, our first step---to address the $k^H$ combinatorial explosion---is to replace the intractable space of $k^H$ concrete execution trajectories with a single, finite abstract trajectory tree. This is achieved through a two-step process: first, a static analysis of the control program's paths, and second, the construction of a temporal tree using the results of that analysis.

\noindent \textbf{1. Static Path-Range Analysis.}
We begin with a one-time static analysis of the control program $g$. We identify all $k$ possible, mutually exclusive execution paths $p_i$ (where $i \in \{1, \ldots, k\}$) within a single control loop. 

Our goal is to find the range of possible control signals that each path $p_i$ can ever produce, over all valid inputs $x$ that trigger that path. For simplicity of notation, we assume $u$ is scalar (our framework supports multi-dimensional signals by finding the ranges for each dimension). We compute the maximum and minimum control signal for the $i$-th path by solving the following optimization problems:
\begin{align*}
    \overline{u}_p = & \max_{x \in \mathbb{R}^n} g^{(p)}(x) \quad \text{ subject to } \quad c^{(p)}(x) = \texttt{True},\\
    \underline{u}_p = & \min_{x \in \mathbb{R}^n} g^{(p)}(x) \quad \text{ subject to } \quad c^{(p)}(x) = \texttt{True}.
\end{align*}

Thanks to modern SMT solvers (e.g., Z3~\cite{z3}), these constrained optimization problems can be efficiently solved for many common program structures. This static analysis yields a finite set $R$ of $k$ abstract control intervals, one for each path:

$$R = \Big\{ [\underline{u}_1, \overline{u}_1], \ldots, [\underline{u}_k, \overline{u}_k] \Big\}.$$

\noindent \textbf{2. Temporal Abstraction Tree Construction.}
This set $R$ of $k$ static control intervals forms the building blocks for our abstract temporal model. We construct an abstract trajectory tree of depth $H$, as illustrated in Figure~\ref{fig:framework_overview}.

\begin{itemize}
    \item The root of the tree represents the set of all possible initial system states at time $t=0$.

    \item Each node in the tree at any depth $t < H$ represents a decision point (i.e., one execution of the control loop).

    \item Each node branches into $k$ children nodes at depth $t+1$.

    \item Each branch (edge) connecting a parent at $t$ to a child at $t+1$ corresponds to one of the $k$ abstract control intervals $[\underline{u}_i, \overline{u}_i]$ from our static analysis.
\end{itemize}

This tree structure is our core abstraction which maps all possible control signals within one branch of code into one node. Our subsequent reachability analysis (described in Section~\ref{sec:nn-reachability}) will propagate sets of states through this tree. At each time step $t$, it will take the set of states at the parent node and compute the $k$ resulting sets of states at the children nodes, one for each abstract control interval. This structure reduces the $k^H$ problem to one that is linear in the horizon $H$ and polynomial in the number of paths $k$ for the reachability analysis.

\subsection{From STL Specification to DeepBern-Net \textit{STL2NN}}

Our core strategy is to perform reachability analysis over the abstract cyber trajectory tree. To do this, we must represent both the system's physical evolution and its logical specification as neural networks, enabling a unified analysis.

\noindent\textbf{$\bullet$ DynamicsNN:}
The first component, \textit{DynamicsNN} denoted as $\nn_f$, is a network that models the physical system's dynamics. It is trained on simulation data to approximate the system's discrete-time behavior: $x_{t+1} = \nn_f(x_t, u_t) + \epsilon_t$, where $x_t$ is the current state, $u_t$ is the control signal from an abstract control interval $[\underline{u}_i, \overline{u}_i]$, and $\epsilon_t$ is a bounded modeling error. For this paper, we assume $\nn_f$ is provided, and critically, that it is represented as a Deep Bernstein Network (DeepBern-Net) to be compatible with our reachability analysis. \b{During forward reachability, this bounded modeling error is handled
conservatively by enlarging the reachable set at each propagation step. As a result, the reachable state sets used in the subsequent analysis already over-approximate the cumulative effect of these one-step errors over the full horizon.}

\noindent\textbf{$\bullet$ STL2NN:}
The second component, \textit{STL2NN} denoted as $\nn_\varphi$, is a network that computes the quantitative robustness $\rho$ of a system trajectory with respect to the STL specification $\varphi$. Crucially, $\nn_\varphi$ is not trained from data. Instead, it is analytically constructed to be functionally equivalent to the given STL formula. The network's architecture directly mirrors the formula's syntactic parse tree: each logical operator ($\wedge, \vee, \neg$) and temporal operator ($G, F, U$) is systematically translated into a corresponding neural layer or recurrent structure. The network's parameters are fixed and pre-determined to precisely implement the mathematical semantics of these operators.

In particular, the quantitative semantics of STL are computed through nested applications of $\min$ and $\max$ operators. As formalized in~\cite{stl}, the Boolean and temporal operators are defined as:

\begin{align*}
\rho(\psi_x,\,\varphi_1 \land \varphi_2,\, t)
    &= \min(\rho(\psi_x,\varphi_1,t),\, \rho(\psi_x,\varphi_2,t)), \\
\rho(\psi_x,\,\varphi_1 \lor \varphi_2,\, t)
    &= \max(\rho(\psi_x,\varphi_1,t),\, \rho(\psi_x,\varphi_2,t)), \\
\rho(\psi_x,\,G_I\varphi_1,\,t)
    &= \min_{t'\in I}\rho(\psi_x,\varphi_1,t+t'), \\
\rho(\psi_x,\,F_I\varphi_1,\,t)
    &= \max_{t'\in I}\rho(\psi_x,\varphi_1,t+t'),
\end{align*}
and the until operator combines both:
\begin{multline*}
\rho(\psi_x,\varphi_1U_I\varphi_2,t) \! \\
    = \! \max_{t'\!\in I}\;
        \! \min \Bigl(
            \rho(\psi_x,\varphi_2,t\!+\!t'),
            \!\min_{\tau \in [0,t'\!]} \rho(\psi_x,\varphi_1,t\!+\!\tau)
        \Bigr).
\end{multline*}
Thus, computing STL robustness reduces entirely to evaluating compositions of $\min$ and $\max$.

\noindent \textbf{$\bullet$ DeepBern-Net Approximation of $\varphi$.}
Prior work reported in~\cite{stl2nn} showed how to use ReLU-based neural networks to represent $\min$ and $\max$ operators and how to combine them to construct ReLU-based neural networks that represent any STL formula $\varphi$. While the ReLU-based representation of $\varphi$ is exact, reachability analysis for ReLU networks yields excessively loose bounds because the ReLU networks form piecewise-linear mappings. Interval or star-set reachability methods must conservatively over-approximate all linear regions that the input set might intersect. As the number of feasible regions grows combinatorially with network depth, this leads to an explosion of a different kind, resulting in increasingly loose and computationally useless reachable-set bounds.

To avoid this piecewise-linear branching, we construct $\nn_\varphi$ using DeepBern-Nets, which requires approximating $\min$ and $\max$ with smooth Bernstein polynomials. The identities:
\begin{align*}
\min(a,b)=\frac{(a+b)-|a-b|}{2},
\max(a,b)=\frac{(a+b)+|a-b|}{2}
\end{align*}
reduce this problem to finding a good approximation for the absolute value function $|x|$ on a bounded input interval $[l,u]$. We construct a Bernstein polynomial for this purpose:

$$B_n(|x|)(x)
=
\sum_{k=0}^{n} c^{|x|}_k \, b_{n,k}^{[l,u]}(x),$$
where the basis polynomials are $b_{n,k}^{[l,u]}(x)$ and the coefficients are chosen as $c^{|x|}_k = |l + (u-l)\frac{k}{n}|$. This allows us to define our approximate operators:\\
\begin{align*}
\mathrm{BernMin}_{n}(a,b)
&=
\frac{(a+b) - B_n(|x|)(a-b)}{2}, \\
\mathrm{BernMax}_{n}(a,b)
&=
\frac{(a+b) + B_n(|x|)(a-b)}{2}.
\end{align*}
As we show in Section~\ref{sec:experiment}, the resulting $\mathrm{BernMin}$ and $\mathrm{BernMax}$ operators are global polynomials in $(a,b)$ expressed directly in Bernstein form. They introduce no piecewise structure. As a result, their reachable bounds under polynomial-based reachability are obtained simply by taking the minimum and maximum of their Bernstein coefficients. This avoids the combinatorial branching of ReLU methods, allowing our $\nn_\varphi$ to yield significantly tighter and more useful bounds.

By cascading $\nn_f$ and $\nn_\varphi$, we create a composite analysis pipeline where we can efficiently compute provable bounds on the robustness score for entire sets of trajectories, which we detail next.

\subsection{NN Reachability Analysis}
\label{sec:nn-reachability}

A key component of our framework is the neural network reachability analysis, which computes output sets on given input sets. This approach allows us to efficiently analyze the behavior of the system over ranges of inputs rather than individual data points. In this work, we construct such neural networks with \textbf{Bernstein activation functions}~\cite{reachnn, bernnn, DeepBern}.
This choice is critical for our framework, as Bernstein polynomials enable the computation of tight, mathematically rigorous bounds on the neural network's output. This property is essential for our reachability-based pruning strategy, as it allows for more precise classification of trajectories and maximizes the number of pruned, provably safe regions.

In the context of this work, our NNs take initial system state bounds as input and approximate the robustness score bound.

By using this forward reachability technique, we can inspect whether a given input system states $\mathcal{X}$ could yield an output robustness score range covering a safe/unsafe region (i.e., positive/negative robustness region). The output regions can be classified into three categories.
\b{Given a reachable robustness interval $\mathcal{R}_{ \nn_f \circ \nn_\varphi}(\mathcal{X}) = [\tilde{\rho}_{\min}, \tilde{\rho}_{\max}]$, we conservatively account for the approximation induced by replacing exact $\min\!/\!\max$ operators in STL2NN with $\mathrm{BernMin}_n/\mathrm{BernMax}_n$. Let $\bar{\epsilon}_{\varphi}$ denote an upper bound on this approximation error, and define the conservative robustness bound as:
$
\rho_{\min} := \tilde{\rho}_{\min} - \bar{\epsilon}_{\varphi},
~~
\rho_{\max} := \tilde{\rho}_{\max} + \bar{\epsilon}_{\varphi}.
$
Accordingly, we use:
\[
R_{NN_f \circ NN_\varphi}(X) = [\rho_{\min}, \rho_{\max}]
\]
for the following classification.}

\begin{itemize}
    \item \textbf{\g{SAFE}}: If the entire output interval lies above zero, i.e., $\rho_{\min} > 0,$ then all trajectories within $\mathcal{X}$ are predicted to satisfy the STL specification.
    \item \textbf{\r{UNSAFE}}: If the entire output interval lies below zero, i.e., $\rho_{\max} < 0,$ then all trajectories within $\mathcal{X}$ are predicted to violate the STL specification.
    \item \textbf{\y{UNCERTAIN}}: If the output interval spans both negative and positive values, i.e., $\rho_{\min} < 0 < \rho_{\max},$ then the region $\mathcal{X}$ may contain both safe and unsafe trajectories, and further investigation is required.
\end{itemize}

In addition to these three classifications, the \textit{DynamicsNN} can also determine \textbf{UNREACHABLE} states in terms of the system dynamics because this NN is supposed to surrogate the system behavior and hence trained to generate the state sequences that are only kinematically feasible. 

\subsection{System Falsification}
After the reachability-guided pruning phase, the \toolname~framework is left with a significantly reduced set of cyber trajectories categorized as \y{UNCERTAIN}. These are the trajectories for which our neural network-based analysis could not definitively prove safety or violation. The final step is to apply a system falsification engine to this set of uncertain trajectories to find concrete cyber-kinetic vulnerabilities.

System falsification can be formulated as an optimization problem. Given a safety requirement specified by an STL formula $\varphi$, the goal is to find an initial physical state $x_0$ and a sequence of input signals $\{s_t\}_{t=0}^{H-1}$ that cause a safety violation of the corresponding physical trajectory $\psi_x$. The falsification engine searches for a combination of $x_0$ and $\{s_t\}$ that minimizes the STL robustness score $\rho(\psi_x, \varphi)$, while ensuring the input signals adhere to the constraints of the chosen cyber trajectory.

Formally, for the set of all uncertain cyber trajectories $\Psi_{uncertain}$, the falsification problem is:
\begin{align*}
\label{eq:falsification}
& \min_{x_0 \in \mathcal{X}_0, \{s_t\}_{t=0}^{H-1}} \rho(\psi_x(x_0, \{s_t\}), \varphi) \\
\text{s.t.} & \quad  c^{(p_t)}(x_t + s_t) = \texttt{True}, \quad \forall t \in [0, H-1)
\end{align*}
where $\psi_x(x_0, \{s_t\})$ is the physical state trajectory $\{x_t\}_{t=0}^{H}$ generated by the dynamics $x_{t+1} = f(x_t, g(h(x_t) + s_t))$ starting from $x_0$. The constraint $c^{(p_t)}(\cdot)$ ensures that the inputs at each step are valid for the specific program path $p_t$ of the cyber trajectory $\psi$.
A solution to this optimization problem with a negative robustness score ($\rho < 0$) constitutes a concrete counterexample, revealing a cyber-kinetic vulnerability.

Falsification engines, such as \emph{S-TaLiRo}~\cite{s-taliro}, typically employ stochastic optimization algorithms (e.g., simulated annealing, genetic algorithms) to solve this non-convex optimization problem. These simulations are computationally expensive. The core benefit of \toolname~is that it invokes this costly falsification process only after drastically reducing the search space. By focusing only on the \y{UNCERTAIN} cyber trajectories—those most likely to contain vulnerabilities—\toolname~significantly improves the overall efficiency and scalability of the vulnerability detection process.

\subsection{Example Run of \toolname}
\begin{figure}[H]
    \centering
    \includegraphics[width=\linewidth]{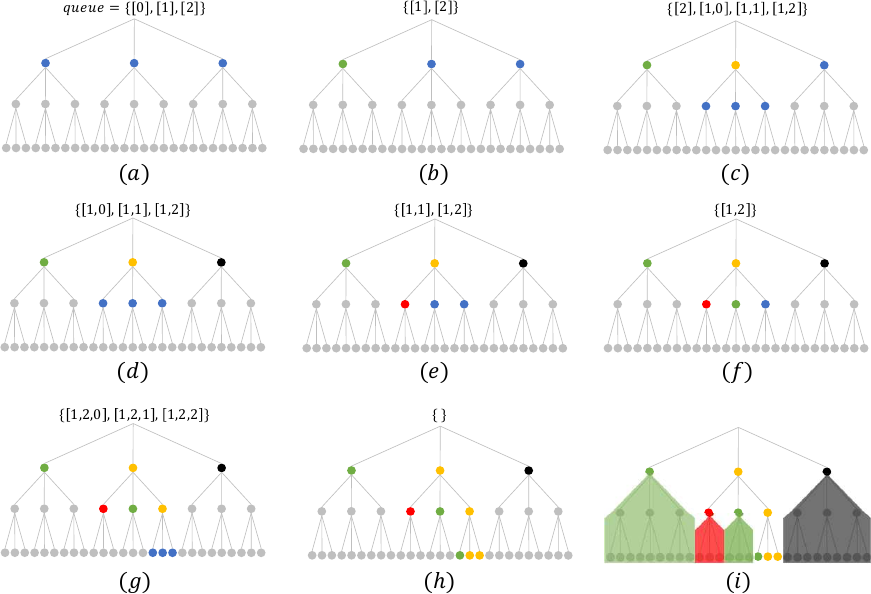}
    \caption{Example Run of \toolname}
    \label{fig:ramponn_tree}
\end{figure}

    We summarize one iteration of \toolname in Figure~\ref{fig:ramponn_tree}.
    (a) The root node is expanded, and its three children $[0], [1], [2]$ are enqueued for exploration. (b) Node $[0]$ is dequeued, the reachability was performed, and pruned as safe $(\rho_{min}\!>\!0)$, leaving $[1], [2]$ in the queue. (c) Node $[1]$ is dequeued and marked uncertain $(\rho_{min} \! \le \! 0 \! \le \! \rho_{max})$, so its children $[1,0], [1,1], [1,2]$ are enqueued; node $[2]$ remains. (d) Node $[1,0]$ is dequeued and marked safe; remaining queue: $[1,1], [1,2]$. (e) Node $[1,1]$ is dequeued and marked safe. (f) Node $[1,2]$ is dequeued and marked uncertain; its children $[1,2,0], [1,2,1], [1,2,2]$ are added. (g) All leaf nodes in the queue are also processed, and (h) added to the corresponding sets. (i) Final sets are highlighted: \g{SAFE}, \r{UNSAFE}, \y{UNCERTAIN}, or \textbf{UNREACHABLE}. This \toolname flow is summarized in Algorithm~\ref{alg:ramponn}.

\subsection{Correctness of Algorithm~\ref{alg:ramponn}}
The soundness and completeness of Algorithm~\ref{alg:ramponn} rely on the soundness and completeness of the falsification engine and the NN reachability analysis. This follows from the fact that \toolname uses sound abstraction of the cyber trajectories and relies on the NN reachability and falsification engine to discard abstract trajectories when no abstract physical-level trajectory violates the safety requirements.
While tree pruning based on NN reachability does not affect the soundness or completeness of \toolname\!\!, off-the-shelf falsification engines rely on stochastic optimization to reason about non-convex constraints. These engines are sound but not complete; therefore, \toolname is also sound but not complete.

\begin{algorithm}[t]
\caption{\toolname}
\label{alg:ramponn}
\begin{algorithmic}[1]
\REQUIRE \textit{DynamicsNN} $\mathcal{N\!N}_{f}$, \textit{STL2NN} $\mathcal{N\!N}_\varphi$, Control Program $g$, Time Horizon $H$
\ENSURE Sets of cyber trajectories \\ $\Psi_{safe}, \Psi_{unsafe}, \Psi_{uncertain}, \Psi_{unreach}$

\STATE Initialize an empty queue: $Q \gets \emptyset$
\STATE Initialize sets: $\Psi_{safe}, \Psi_{unsafe}, \Psi_{uncertain}, \Psi_{unreach} \gets \emptyset$
\STATE $\psi_\emptyset \gets \text{GetRootNode}(g)$
\STATE $Q.\text{enqueue}(\psi_\emptyset.\text{children})$ \COMMENT{Enqueue branches}

\WHILE{$Q$ is not empty}
    \STATE $\psi_l \gets Q.\text{dequeue}()$ \COMMENT{Get a prefix of a cyber trajectory}
    \IF{$\psi_l$ has been visited}
        \STATE \textbf{continue}
    \ENDIF
    \STATE Mark $\psi_l$ as visited

    \STATE $X_{l} \gets \text{GetStateBoundsForPrefix}(\psi_l, g)$ \COMMENT{Get physical state bounds for the prefix}
    \STATE $h \gets H - l$ \COMMENT{Remaining steps}
    
    \STATE $\hat{X}_{seq} \gets \text{MultiStepReachability}(\mathcal{N\!N}_{f}, X_{l}, h)$ \COMMENT{Predict reachable physical states}
    
    \STATE $[\rho_{min}, \rho_{max}] \gets \text{Reachability}(\mathcal{N\!N}_\varphi, \hat{X}_{seq})$

    \IF{$\rho_{min} > 0$}
        \STATE $\Psi_{safe} \gets \Psi_{safe} \cup \{\psi_l\}$
        \STATE \textbf{continue} \COMMENT{Prune the entire subtree as \textbf{\underline{\g{SAFE}}}}
    \ELSIF{$\rho_{max} < 0$}
        \STATE $\Psi_{unsafe} \gets \Psi_{unsafe} \cup \{\psi_l\}$
        \STATE \textbf{continue} \COMMENT{Prune the entire subtree as \textbf{\underline{\r{UNSAFE}}}}
    \ELSE
        \STATE \textbf{\underline{\y{UNCERTAIN}}}
        \IF{$\text{IsLeafNode}(\psi_l)$}
            \STATE $\Psi_{uncertain} \gets \Psi_{uncertain} \cup \{\psi_l\}$ \COMMENT{Add full trajectory to candidates}
        \ELSE
            \STATE $B_{reachable} \gets \text{FindReachableBranches}(\hat{X}_{seq}, g)$
            \FORALL{children $\psi_{l+1}$ of $\psi_l$}
                \IF{branch leading to $\psi_{l+1}$ is in $B_{reachable}$}
                    \STATE $Q.\text{enqueue}(\psi_{l+1})$ \COMMENT{Explore promising children}
                \ELSE
                    \STATE $\Psi_{unreach} \gets \Psi_{unreach} \cup \{\psi_{l+1}\}$ \COMMENT{Prune the entire subtree as \textbf{\underline{UNREACHABLE}}}
                \ENDIF
            \ENDFOR
        \ENDIF
    \ENDIF
\ENDWHILE

\RETURN $\Psi_{safe}, \Psi_{unsafe}, \Psi_{uncertain}, \Psi_{unreach}$
\end{algorithmic}
\end{algorithm}

\section{Experiment and Evaluation}
\label{sec:experiment}

We conduct a series of experiments to evaluate the effectiveness and efficiency of the \toolname~framework. Our evaluation is designed to answer three primary questions:

\noindent $\bullet$ How does the choice of network architecture (DeepBern-Nets vs. ReLU) impact model expressiveness and the tightness of reachability analysis for \textit{DynamicsNN} and \textit{STL2NN}?

\noindent $\bullet$ How does \toolname~scale in terms of analysis time and vulnerability detection capability as the system complexity and time horizon $H$ increase, compared to baseline approaches?

\noindent $\bullet$ Can \toolname~be effectively used in an iterative development loop to guide a developer in finding and fixing cyber-kinetic vulnerabilities?

\subsection{Experimental Setup}

\subsubsection{\underline{Benchmarks}}
We use two distinct CPS benchmarks to test our framework's performance across different complexities:

\noindent $\bullet$ \textbf{PLC Water Tank:} A 1-state model representing a water tank controlled by a Programmable Logic Controller (PLC). The control logic is simple, but its interaction with the continuous water-level dynamics can lead to non-trivial vulnerabilities (e.g., overflow or underflow). 
The controller manipulates four inputs: an input-valve command $u_{\mathrm{in}}^{\mathrm{bool}} \in \{0,1\}$, an output-valve command $u_{\mathrm{out}}^{\mathrm{bool}} \in \{0,1\}$, a real-valued inflow rate $q_{\mathrm{in}}$, and a real-valued outflow rate $q_{\mathrm{out}}$. The Boolean valve commands induce hybrid switching, allowing the PLC code to produce complex cyber trajectories (e.g., repeated filling/emptying cycles). The safety requirement for this benchmark is $G_{[0,30]} \bigl( \text{TankHeight} \le 8 \bigr),$ meaning ``the water level must not exceed~8 for 30\,s of simulation time.'' Below is the control software used to control the Water Tank.\\

\begin{lstlisting}[caption={Control Program for Water Tank Model}, label={lst:control_program}][H]
double watertank_control(double TankHeight) {
    double FullHeight = 10.0;
    double EmptyHeight = 5.0;
    double MidHeight = 7.0;

    if (TankHeight <= EmptyHeight) {
        InValve = true;
        OutValve = false;
        InValveRate = 1.0;
        OutValveRate = 0.0;
    } else if (TankHeight >= FullHeight) {
        InValve = false;
        OutValve = true;
        InValveRate = 0.0;
        OutValveRate = 1.0;
    } else if (TankHeight > EmptyHeight && TankHeight < MidHeight) { 
        InValve = true;
        OutValve = false;
        InValveRate = (MidHeight - TankHeight) / (MidHeight - EmptyHeight);
        OutValveRate = 0.0;
    } else {
        InValve = false;
        OutValve = true;
        InValveRate = 0.0;
        OutValveRate = (TankHeight - MidHeight) / (FullHeight - MidHeight);
    }
    return InValve, OutValve, InValveRate, OutValveRate;
}
\end{lstlisting}

\noindent $\bullet$ \textbf{Switched PID Automotive Engine:} A higher-dimensional model (e.g., $>10$ states) of an automotive engine. The control code implements a switched PID controller, which introduces significant branching complexity ($k$) in the cyber trajectory tree. The plant receives a 3-dimensional input consisting of engine speed (RPM), vehicle speed, and throttle command, and outputs the next-step RPM and speed (2-dimensional output). The STL safety specification is $G_{[0,30]} \bigl( \text{Speed} < 100 \;\lor\; \text{RPM} < 4300 \bigr),$ requiring the vehicle speed to stay below 100\,mph or the engine RPM to remain below~4300.

\begin{lstlisting}[caption={Control Program for Automotive Engine Model}, label={lst:control_program_ae}]
double control(double RPM, double Speed){
  if (RPM > 3300){
    if (Speed > 80){
      Throttle = -RPM*0.002 - Speed*1.1 + 183.0;
    } else {
      Throttle = - RPM*0.001 + Speed*0.6 + 19.0;
    }
  } else {
    if (Speed > 80){
      Throttle = RPM*0.001 - Speed*1.7 + 216.0;
    } else {
      Throttle = - RPM*0.001 - Speed*0.6 + 139.0;
    }
  }
  return Throttle;
}
\end{lstlisting}

\begin{table*}[t]
\centering
\caption{Test Loss and Tightness of the NN reachable sets for ReLU and DeepBern-Nets using Water Tank Control System}
\vspace{-2mm}
\setlength{\tabcolsep}{4pt}
\small
\begin{tabular}{l c c c c c c}
\toprule
\textbf{NN Activation} & \textbf{Layers} & \textbf{Bern-Polynomial} & \textbf{Number of} & \textbf{MSE} & \textbf{Reachable Set} & \textbf{Tightness [\%]} \\
& & \textbf{Degree} & \textbf{Params} & & \textbf{Volume} & \\
\midrule
\multirow{5}{*}{\textbf{ReLU-Based NN}}
 & [128,128,128]               & - & 33,921    & 0.6161 & 74.9016    & 323.59 \\
 & [256,256,256]               & - & 133,377   & 0.6629 & 153.0460   & 765.52 \\
 & [512,512,512]               & - & 528,897   & 0.6451 & 376.1116   & 2027.01 \\
 & [512,512,512,512]           & - & 791,553   & 1.2853 & 1243.9788  & 6935.02 \\
 & [1024,1024,1024,1024]       & - & 3,155,969 & 0.6946 & 3894.9106  & 21926.74 \\
\midrule
\multirow{6}{*}{\textbf{DeepBern-Net}}
 & [128,128,128]               & 3 & 35,457    & 0.0254 & 33.1993    & 87.75 \\
 & [256,256,256]               & 3 & 136,449   & 0.0251 & 34.8529    & 97.10 \\
 & [512,512,512]               & 3 & 535,041   & 0.0234 & 46.0599    & 160.48 \\
 & [512,512,512]               & 4 & 536,577   & 0.0233 & 44.7154    & 152.88 \\
 & [1024,1024,1024]            & 4 & 2,121,729 & 0.0242 & 47.9205    & 171.00 \\
 & [1024,1024,1024,1024]       & 4 & 3,176,449 & 0.0249 & 44.2336    & 150.15 \\
\bottomrule
\end{tabular}
\label{tab:watertank_small_domain_full}
\end{table*}

\begin{table*}[t]
\centering
\caption{Test Loss and Tightness of the NN reachable sets for ReLU and DeepBern-Nets using Engine Control System}
\vspace{-2mm}
\setlength{\tabcolsep}{4pt}
\small
\begin{tabular}{l c c c c c c}
\toprule
\textbf{NN Activation} & \textbf{Layers} & \textbf{Bern-Polynomial} & \textbf{Number of} & \textbf{MSE} & \textbf{Reachable Set} & \textbf{Tightness[\%]} \\
& & \textbf{Degree} & \textbf{Params} & & \textbf{Volume} & \\
\midrule
\multirow{7}{*}{\textbf{ReLU-Based NN}}
 & [256,256,256]                    & - & 133,122    & 0.3194 & 2200455.5      & 293.45 \\
 & [512,512,512]                    & - & 528,386    & 0.2444 & 6695192.5      & 1097.12 \\
 & [512,512,512,512]                & - & 791,042    & 0.5829 & 14187949.0     & 2436.85 \\
 & [1024,1024,1024,1024]            & - & 3,154,946  & 0.2834 & 86667400.0     & 15396.43 \\
 & [2048,2048,2048,2048]            & - & 12,601,346 & 0.1434 & 4737901568.0   & 847052.99 \\
 & [2048,2048,2048,2048,2048]       & - & 16,797,698 & 0.4374 & 10459015168.0  & 1870007.65 \\
 & [2048,2048,2048,2048,2048,2048]  & - & 20,994,050 & 0.9665 & 182017802240.0 & 32545305.07 \\
\midrule
\multirow{8}{*}{\textbf{DeepBern-Net}}
 & [256,256,256]                    & 3 & 136,194    & 0.2478 & 1315818.75     & 135.27 \\
 & [256,256,256]                    & 4 & 136,962    & 0.2432 & 1885965.25     & 237.22 \\
 & [512,512,512]                    & 4 & 536,066    & 0.2412 & 3354818.0      & 499.85 \\
 & [512,512,512,512]                & 4 & 801,282    & 0.2416 & 1760191.875    & 214.73 \\
 & [1024,1024,1024,1024]            & 4 & 3,175,426  & 0.2390 & 2820024.25     & 404.23 \\
 & [2048,2048,2048,2048]            & 4 & 12,642,306 & 0.2388 & 4300155.5      & 668.88 \\
 & [2048,2048,2048,2048]            & 8 & 12,675,074 & 0.0014 & 10016671.0     & 1691.02 \\
 & [2048,2048,2048,2048,2048]       & 8 & 16,889,858 & 0.0014 & 11190077.0     & 1900.82 \\
\bottomrule
\end{tabular}
\normalsize
\label{tab:engine_full_results}
\vspace{-5mm}
\end{table*}

\subsubsection{\underline{Baseline Approaches}}
We compare \toolname~against three representative baseline methods:

\noindent $\bullet$ \textbf{Standalone/Unguided Falsification Engines (S-TaLiRo):} A state-of-the-art, unguided falsification engine (e.g., based on S-TaLiRo~\cite{s-taliro}) that uses optimization to directly search for specification violations in every possible abstract cyber trajectory the full-system model.

\noindent $\bullet$ \textbf{Counter-Example Guided Abstraction Refinement Based Falsification (Rampo+S-TaLiRo):} Prior work~\cite{kohei} proposed the use of iterative abstraction refinement techniques to guide Falsification engines (e.g., S-TaLiRo) to better traverse the tree of cyber trajectories. Such methods do not use neural network reachability and are based on tree search only.

\noindent $\bullet$ \textbf{ReLU-Based Guided Falsification (ReLU-based \toolname):} A re-implementation of our \toolname~framework, but using standard ReLU-based neural networks for \textit{DynamicsNN} and an equivalent ReLU-based \textit{STL2NN} (proposed in~\cite{stl2nn}), analyzed with auto\_LiRPA \cite{autolirpa}; a standard interval-based ReLU reachability tool using an algorithm named \textbf{$\boldsymbol{\alpha}$-\emph{CROWN}}~\cite{xu2021fast}.

\subsubsection{\underline{Implementation Details}}
All experiments are conducted on an Ubuntu 20.04.6 LTS server equipped with dual Intel Xeon E5-2650 v4 CPUs (48 cores, 96 threads), 256 GB RAM, and 8× NVIDIA GeForce RTX 2080 Ti GPUs. \textit{DynamicsNN} models are trained using simulation data from Simulink models of the benchmarks. Our reachability analysis and abstract trajectory tree exploration are implemented in Python, leveraging \emph{angr}~\cite{angr} and \emph{Z3}~\cite{z3} SMT solver for static path analysis. For the falsification engine, we used \emph{S-TaLiRo} to perform the system falsification and \emph{DeepBern-Nets}~\cite{DeepBern} for the construction of \textit{DynamicsNN} and \textit{STL2NN}.

\subsection{Experiment 1: Efficacy of DeepBern-Nets vs. ReLU}

\noindent $\bullet$ \textbf{Objective.}
This experiment validates our core architectural choice. We isolate the \textit{DynamicsNN} component to compare the approximation power (expressiveness) and reachability tightness of DeepBern-Nets against traditional ReLU networks.

\noindent $\bullet$ \textbf{Methodology.}
For both the Water Tank and Automotive Engine benchmarks, we train several \textit{DynamicsNN} models of varying sizes (both width and depth) using identical training datasets.
To ensure a fair and rigorous comparison, our evaluation of ReLU networks was exhaustive. For the ReLU models, we performed an extensive grid search over hyperparameters (e.g., learning rate, initial weights) and selected only the best-performing architecture we could find for our comparison. 
We compare (i) the final empirical test loss (i.e., MSE) to measure expressiveness and (ii) the tightness of the reachable set bounds for a given input set. For tightness, we compute a ``ground truth'' reachable set by performing a uniform random sampling (e.g., $10^7$ samples) and compare the volume of the over-approximated reachable sets computed by the DeepBern and ReLU analysis tools. That is, we use the following  $\mathrm{Tightness}$ score:
\[
\mathrm{Tightness} [\%]
=
\frac{ V_{\mathrm{reach}} - V_{\mathrm{sampling}} }{ V_{\mathrm{sampling}} } \times 100,
\]
where $V_{\mathrm{sampling}}$ and $V_{\mathrm{reach}}$ are the volume of the hypercubes computed using NN reachability and sampling, respectively. A value of \(0\) indicates perfect agreement, while positive values quantify 
the degree of over-approximation.  
Negative values indicate that the predicted interval is smaller than the observed box, 
corresponding to an unsound approximation.

\noindent $\bullet$ \textbf{Results.}
As shown in Table~\ref{tab:watertank_small_domain_full} (for the Water Tank benchmark), a compact DeepBern-Net consistently achieves a lower approximation loss than a much larger ReLU network. For example, the smallest DeepBern-Net with an architecture of [128,128,128] achieves an order of magnitude lower validation loss compared to all the ReLU-based neural networks including those with more than 10000x trainable parameters (e.g., [1024,1024,1024,1024]). A similar conclusion can also be drawn from Table~\ref{tab:engine_full_results} for the Engine Model.

Furthermore, Table~\ref{tab:watertank_small_domain_full} and Table~\ref{tab:engine_full_results} also show that DeepBern-Net reachability bounds are orders of magnitude tighter. For example, in the Engine Control model, the reachable set computed for the largest DeepBern-Net has a volume that is just 19.0082\% larger than the ground truth, whereas the ReLU-based analysis produces a bound that is over 1000x larger. This is because standard reachability tools must conservatively account for the combinatorial splitting of piecewise-linear regions in ReLU networks (used in the B-ReLU baseline), leading to unusable, explosive bounds. 

\noindent $\bullet$ \textbf{Conclusion.}
DeepBern-Nets are superior for this task in two critical ways: (i) they provide more accurate approximations of the physical dynamics, and (ii) their mathematical structure permits reachability analysis that is orders of magnitude tighter. The looseness of ReLU bounds would lead to spurious \y{UNCERTAIN} classifications, rendering the pruning in \toolname~ineffective.

\subsection{Experiment 2: Scalability Evaluation}

\noindent $\bullet$ \textbf{Objective.}
This experiment evaluates the end-to-end performance and scalability of the \toolname framework against baselines by analyzing the benchmarks with an increasing time horizon $H$.

\noindent $\bullet$ \textbf{Methodology.}
We run \toolname\!\!, Rampo~+~S-TaLiRo, S-TaLiRo, and ReLU-based \toolname on both the Water Tank and Automotive Engine benchmarks. We vary the time horizon $H$ from a shallow $H=2$ to a deep $H=10$, which increases the concrete trajectory space from $k^{2}$ to $k^{10}$. We measure (i) total execution time and (ii) the number of unique specification-violating trajectories (vulnerabilities) found. We set a timeout of 3 hours for each run.

\begin{figure}[H]
    \vspace{-5mm}
    \centering
    \includegraphics[width=0.99\linewidth]{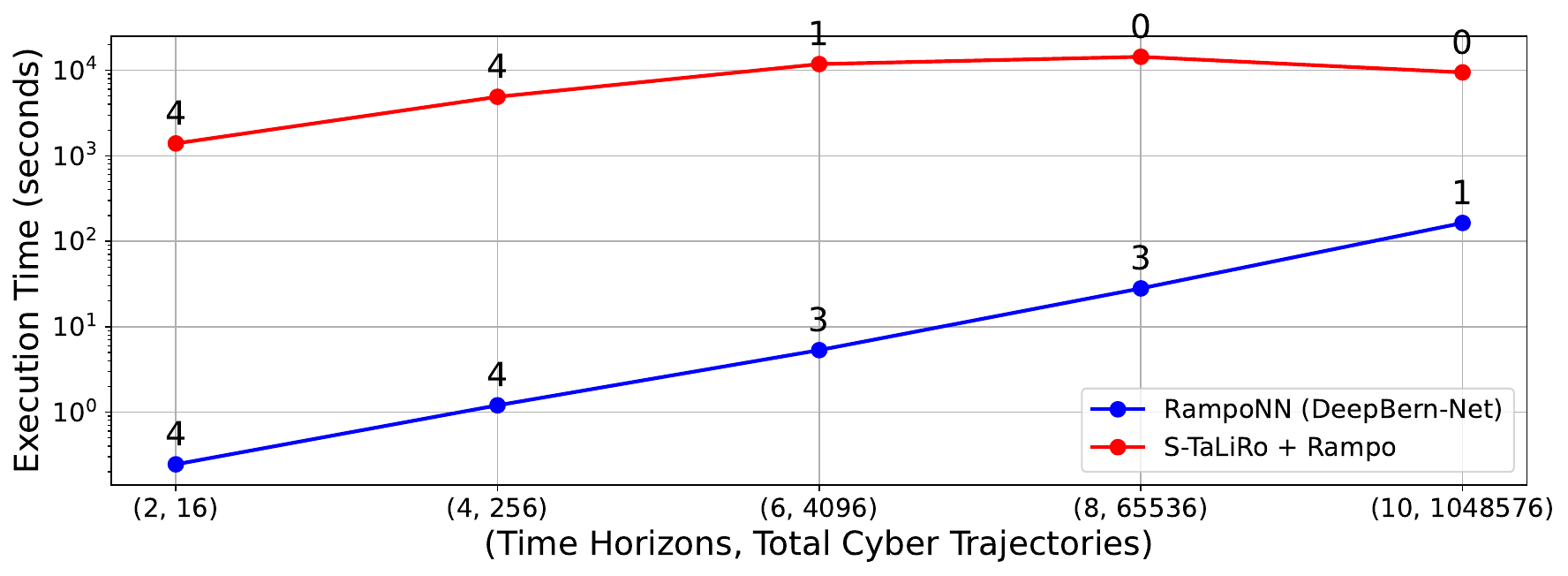} \\
    \includegraphics[width=0.99\linewidth]{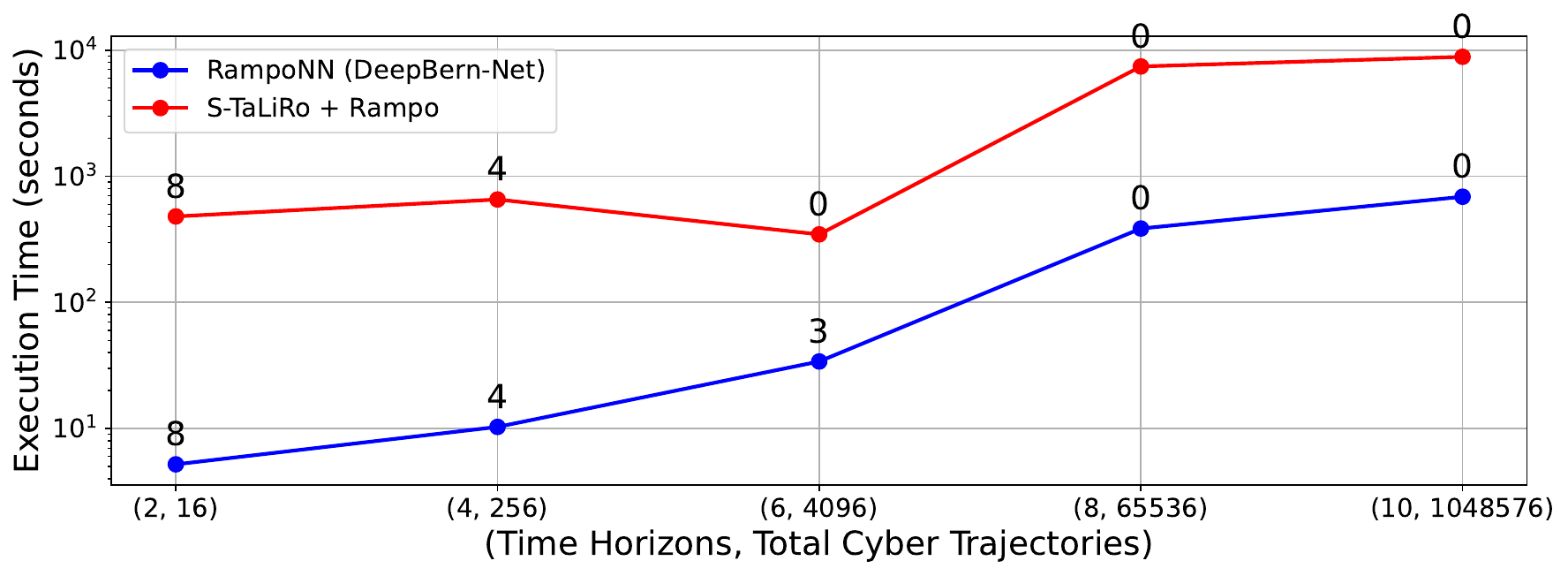} 
    \vspace{-5mm}
    \caption{Execution Time and number of identified vulnerabilities for different time horizons (Top) Water Tank model (Bottom) Automotive Engine model.}
    \label{fig:executiontime}
\end{figure}

\noindent $\bullet$ \textbf{Results.}
The results are summarized in Figure~\ref{fig:executiontime} and as follows:
\begin{itemize}
    \item \underline{Unguided falsification (S-TaliRo)}: Fails to find any deep-nested vulnerabilities due to timeout. This shows the challenge of exploring large abstract cyber-trees.

    \item \underline{ReLU-based \toolname}
    Due to the loose bounds established in Experiment 1, it classifies almost all abstract trajectories as \y{UNCERTAIN}. It prunes almost nothing and hence the falsification engine is challenged by exploring all possible abstract cyber trajectories which again times out and fail to identify any vulnerabilities. 

    \item \underline{Rampo~+~S-TaLiRo} was able to traverse the entire abstract cyber trajectory tree and identify a few vulnerabilities. Nevertheless, it requires a high number of calls to the falsification engine (S-TaLiRo) which in itself is a stochastic engine. The higher the number this engine is used, the higher the chance it may miss some vulnerabilities. This is reflected in a lower number of identified vulnerabilities.

    \item \underline{\toolname (ours)} consistently outperforms all other tools. Its execution time scales more favorably with $H$, as the reachability analysis efficiently prunes vast swaths of the \g{SAFE} abstract trajectory tree. It successfully finds the highest number of vulnerabilities and is the only tool to discover the deep-nested vulnerabilities in the automotive engine model at $H=10$. 
    \b{Although a higher continuous-state dimension can increase the difficulty of each downstream falsification call through more challenging simulation and optimization, this effect alone does not determine the end-to-end runtime. In practice, the overall cost is also strongly shaped by how many prefixes remain \y{UNCERTAIN} after reachability-based pruning and must be further examined by the falsification engine.}
\end{itemize}

\noindent $\bullet$ \textbf{Conclusion.}
\toolname is the only framework that is both scalable and effective. Unguided falsification does not scale, and standard ReLU-based reachability is not tight enough to be useful for pruning. \toolname\!'s DeepBern-Net analysis provides the ``best of both worlds'': a sound, tight pruning that makes the subsequent falsification problem tractable.

\subsection{Experiment 3: Case Study on Guided Vulnerability Remediation}

\noindent $\bullet$ \textbf{Objective.} This experiment demonstrates \toolname\!'s practical utility as a tool to guide a developer through an iterative ``detect-and-fix'' cycle.

\noindent $\bullet$ \textbf{Methodology.}
We use the Automotive Engine benchmark and begin with a simple, known-buggy PID controller (V1). We run \toolname to find vulnerabilities. We then ``fix'' the control code to create a more complex, presumably safer version (V2). We repeat this process until \toolname can no longer find any vulnerabilities.

\noindent $\bullet$ \textbf{Results.}
Our iterative development process proceeded as follows:

\begin{itemize}
    \item \underline{Code V1 (Simple PID):} The software developer starts with a simple code with no branches:
\begin{lstlisting}
double control(double RPM, double Speed){
    Throttle = - RPM*0.001 - Speed*0.6 + 139.0;
    return Throttle;
}
\end{lstlisting}
    ~\toolname is executed. It rapidly analyzes the abstract tree and guides the falsifier to a critical vulnerability in 0.667 minutes, where a specific sequence of control inputs leads to bypassing the safe Speed limits as shown in the physical trajectories below. 

    \begin{minipage}[t]{0.45\textwidth} 
        \includegraphics[width=0.99\linewidth]{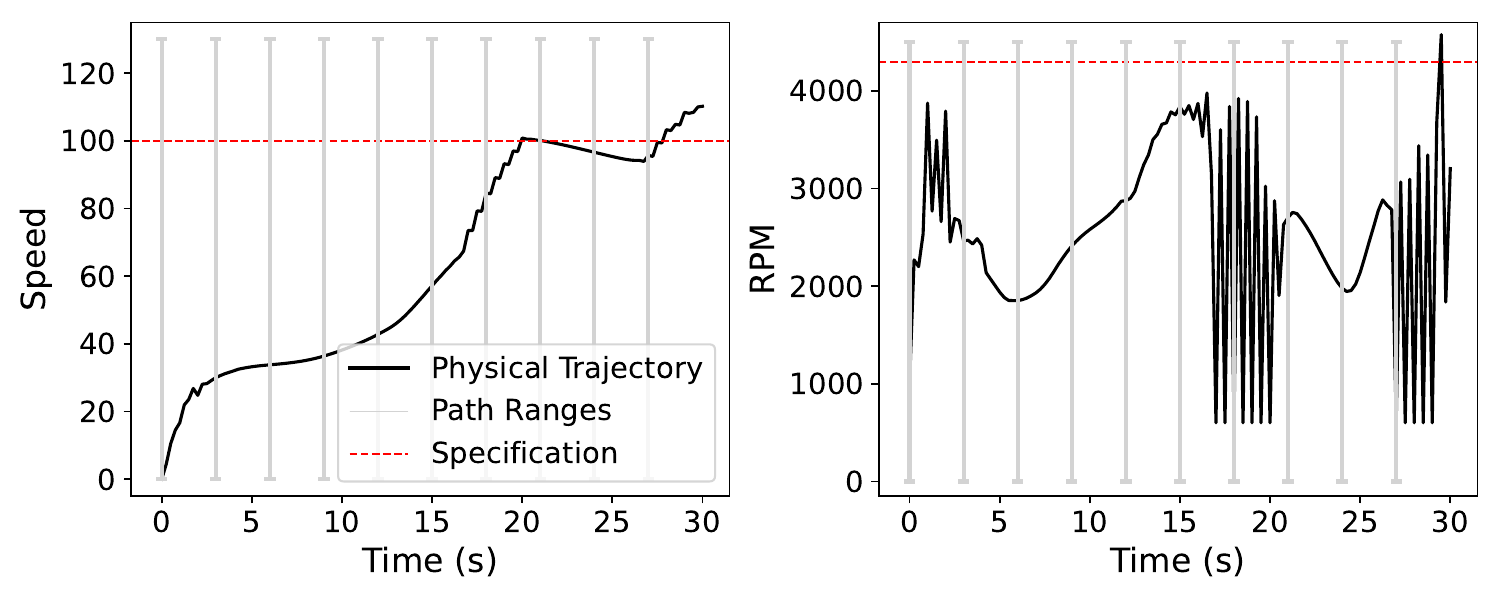}
    \end{minipage}
    The corresponding cyber trajectory is [0, 0, 0, 0, 0, 0, 0, 0, 0, 0], which indicates that branch number 0 (the only possible path in this code) is the one executed across all the time instants leading to this violation of the safety requirements.

    \item \underline{Code V2 (V1 + Safety Guard):}
    The developer adds a new control path (a ``safety mode'') to the code to catch the V1 bug and prevent the speed from overshooting:
    \begin{lstlisting}
double control(double RPM, double Speed){
    if (Speed > 80){
      Throttle = -RPM*0.002 - Speed*1.1 + 183.0;
    } else {
      Throttle = - RPM*0.001 - Speed*0.6 + 139.0;
    }
  return Throttle;
}
\end{lstlisting}
    The code is now more complex. We re-run~\toolname which took 1.07 minutes to analyze V2 code. The tool correctly prunes the V1 trajectory as \g{SAFE}. However, it now guides the falsifier to a new, more subtle vulnerability in the transition logic to the safety mode itself  (shown in the Figure below) corresponding to the cyber trajectory [0,0,0,0,0,1,1,1,1,1].
    \begin{minipage}[t]{0.45\textwidth}
        \includegraphics[width=0.99\linewidth]{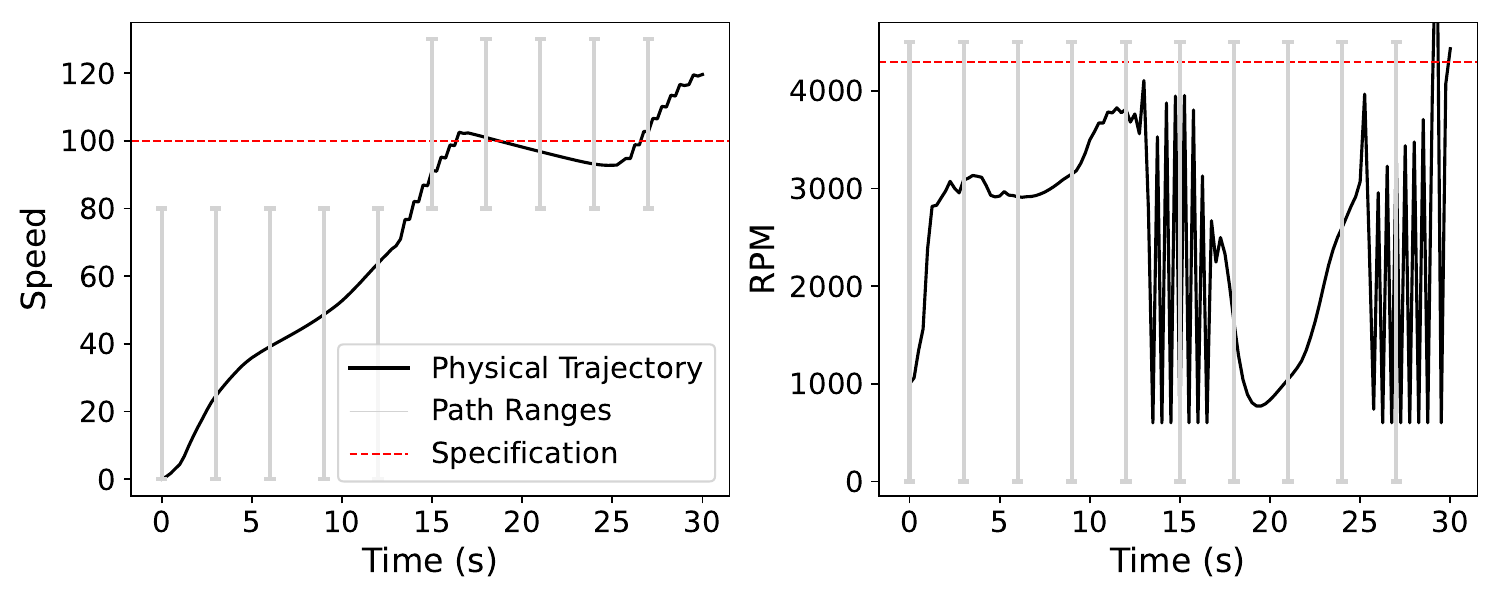}
    \end{minipage}
    \vspace{-6mm}
    \item \underline{Code V3 (V2 + Fixed Guard Logic):} The developer fixes the guard logic leading to the code in Listing~\ref{lst:control_program_ae}.

    We re-run \toolname which takes 17 minutes. 
    The tool analyzes the entire $k^H$ abstract trajectory tree and classifies all subtrees as \g{SAFE}. The falsification engine, guided to the few remaining \y{UNCERTAIN} paths, confirms that no vulnerabilities are found.
\end{itemize}

\section{Conclusion}
\toolname is highly effective as a practical development tool. It not only finds vulnerabilities but also provides targeted guidance for remediation. It successfully ``confirms'' the developer's fix for the V1 bug while simultaneously—and correctly—identifying a new, second-order bug introduced by the fix itself, preventing a false sense of security.
\vspace{-3mm}

\newpage

\bibliographystyle{IEEEtran}
\bibliography{bibliography}

@inproceedings{
xu2021fast,
title={Fast and Complete: Enabling Complete Neural Network Verification with Rapid and Massively Parallel Incomplete Verifiers},
author={Kaidi Xu and Huan Zhang and Shiqi Wang and Yihan Wang and Suman Jana and Xue Lin and Cho-Jui Hsieh},
booktitle={International Conference on Learning Representations},
year={2021},
url={https://openreview.net/forum?id=nVZtXBI6LNn}
}

@INPROCEEDINGS{9797531,
  author={Sheikhi, Sanaz and Kim, Edward and Duggirala, Parasara Sridhar and Bak, Stanley},
  booktitle={2022 ACM/IEEE 13th International Conference on Cyber-Physical Systems (ICCPS)}, 
  title={Coverage-Guided Fuzz Testing for Cyber-Physical Systems}, 
  year={2022},
  volume={},
  number={},
  pages={24-33},
  doi={10.1109/ICCPS54341.2022.00009}}

@inproceedings{10.1145/3395363.3397376,
author = {Chen, Yuqi and Xuan, Bohan and Poskitt, Christopher M. and Sun, Jun and Zhang, Fan},
title = {Active fuzzing for testing and securing cyber-physical systems},
year = {2020},
isbn = {9781450380089},
publisher = {Association for Computing Machinery},
address = {New York, NY, USA},
url = {https://doi.org/10.1145/3395363.3397376},
doi = {10.1145/3395363.3397376},
booktitle = {Proceedings of the 29th ACM SIGSOFT International Symposium on Software Testing and Analysis},
pages = {14–26},
numpages = {13},
location = {Virtual Event, USA},
series = {ISSTA 2020}
}

@InProceedings{pgfuzz,
  author    = {Hyungsub Kim and M. Ozgur Ozmen and Antonio Bianchi and Z. Berkay Celik and Dongyan Xu},
  booktitle = {Network and Distributed System Security Symposium (NDSS)},
  title     = {{PGFUZZ: Policy-Guided Fuzzing for Robotic Vehicles}},
  pages = {1-18},
  comment   = {Acceptance Rate: 15.2\%)},
  year      = {2021},
  url = {https://beerkay.github.io/papers/Berkay2021PGFuzzNDSS.pdf}
}

@inproceedings{10.1145/3548606.3560558,
author = {Kim, Seulbae and Liu, Major and Rhee, Junghwan "John" and Jeon, Yuseok and Kwon, Yonghwi and Kim, Chung Hwan},
title = {DriveFuzz: Discovering Autonomous Driving Bugs through Driving Quality-Guided Fuzzing},
year = {2022},
isbn = {9781450394505},
publisher = {Association for Computing Machinery},
address = {New York, NY, USA},
url = {https://doi.org/10.1145/3548606.3560558},
doi = {10.1145/3548606.3560558},
booktitle = {Proceedings of the 2022 ACM SIGSAC Conference on Computer and Communications Security},
pages = {1753–1767},
numpages = {15},
location = {Los Angeles, CA, USA},
series = {CCS '22}
}

@InProceedings{10.1007/978-3-540-31954-2_19,
author="Girard, Antoine",
editor="Morari, Manfred
and Thiele, Lothar",
title="Reachability of Uncertain Linear Systems Using Zonotopes",
booktitle="Hybrid Systems: Computation and Control",
year="2005",
publisher="Springer Berlin Heidelberg",
address="Berlin, Heidelberg",
pages="291--305",
isbn="978-3-540-31954-2"
}

@inproceedings{10.1007/978-3-030-53288-8_27,
author = {Devonport, Alex and Khaled, Mahmoud and Arcak, Murat and Zamani, Majid},
title = {PIRK: Scalable Interval Reachability Analysis for High-Dimensional Nonlinear Systems},
year = {2020},
isbn = {978-3-030-53287-1},
publisher = {Springer-Verlag},
address = {Berlin, Heidelberg},
url = {https://doi.org/10.1007/978-3-030-53288-8\_27},
doi = {10.1007/978-3-030-53288-8\_27},
booktitle = {Computer Aided Verification: 32nd International Conference, CAV 2020, Los Angeles, CA, USA, July 21–24, 2020, Proceedings, Part I},
pages = {556–568},
numpages = {13},
location = {Los Angeles, CA, USA}
}

@inproceedings{10.1145/3049797.3049808,
author = {Bak, Stanley and Duggirala, Parasara Sridhar},
title = {HyLAA: A Tool for Computing Simulation-Equivalent Reachability for Linear Systems},
year = {2017},
isbn = {9781450345903},
publisher = {Association for Computing Machinery},
address = {New York, NY, USA},
url = {https://doi.org/10.1145/3049797.3049808},
doi = {10.1145/3049797.3049808},
booktitle = {Proceedings of the 20th International Conference on Hybrid Systems: Computation and Control},
pages = {173–178},
numpages = {6},
location = {Pittsburgh, Pennsylvania, USA},
series = {HSCC '17}
}

@inproceedings{10.1145/3302504.3311804,
author = {Bogomolov, Sergiy and Forets, Marcelo and Frehse, Goran and Potomkin, Kostiantyn and Schilling, Christian},
title = {JuliaReach: a toolbox for set-based reachability},
year = {2019},
isbn = {9781450362825},
publisher = {Association for Computing Machinery},
address = {New York, NY, USA},
url = {https://doi.org/10.1145/3302504.3311804},
doi = {10.1145/3302504.3311804},
booktitle = {Proceedings of the 22nd ACM International Conference on Hybrid Systems: Computation and Control},
pages = {39–44},
numpages = {6},
location = {Montreal, Quebec, Canada},
series = {HSCC '19}
}

@InProceedings{10.1007/978-3-319-63387-9_5,
author="Katz, Guy
and Barrett, Clark
and Dill, David L.
and Julian, Kyle
and Kochenderfer, Mykel J.",
editor="Majumdar, Rupak
and Kun{\v{c}}ak, Viktor",
title="Reluplex: An Efficient SMT Solver for Verifying Deep Neural Networks",
booktitle="Computer Aided Verification",
year="2017",
publisher="Springer International Publishing",
address="Cham",
pages="97--117",
isbn="978-3-319-63387-9"
}

@InProceedings{10.1007/978-3-030-25540-4_26,
author="Katz, Guy
and Huang, Derek A.
and Ibeling, Duligur
and Julian, Kyle
and Lazarus, Christopher
and Lim, Rachel
and Shah, Parth
and Thakoor, Shantanu
and Wu, Haoze
and Zelji{\'{c}}, Aleksandar
and Dill, David L.
and Kochenderfer, Mykel J.
and Barrett, Clark",
editor="Dillig, Isil
and Tasiran, Serdar",
title="The Marabou Framework for Verification and Analysis of Deep Neural Networks",
booktitle="Computer Aided Verification",
year="2019",
publisher="Springer International Publishing",
address="Cham",
pages="443--452",
isbn="978-3-030-25540-4"
}

@INPROCEEDINGS{8418593,
  author={Gehr, Timon and Mirman, Matthew and Drachsler-Cohen, Dana and Tsankov, Petar and Chaudhuri, Swarat and Vechev, Martin},
  booktitle={2018 IEEE Symposium on Security and Privacy (SP)}, 
  title={AI2: Safety and Robustness Certification of Neural Networks with Abstract Interpretation}, 
  year={2018},
  volume={},
  number={},
  pages={3-18},
  doi={10.1109/SP.2018.00058}}

@InProceedings{10.1007/978-3-319-77935-5_9,
author="Dutta, Souradeep
and Jha, Susmit
and Sankaranarayanan, Sriram
and Tiwari, Ashish",
editor="Dutle, Aaron
and Mu{\~{n}}oz, C{\'e}sar
and Narkawicz, Anthony",
title="Output Range Analysis for Deep Feedforward Neural Networks",
booktitle="NASA Formal Methods",
year="2018",
publisher="Springer International Publishing",
address="Cham",
pages="121--138",
isbn="978-3-319-77935-5"
}

@ARTICLE{8318388,
  author={Xiang, Weiming and Tran, Hoang-Dung and Johnson, Taylor T.},
  journal={IEEE Transactions on Neural Networks and Learning Systems}, 
  title={Output Reachable Set Estimation and Verification for Multilayer Neural Networks}, 
  year={2018},
  volume={29},
  number={11},
  pages={5777-5783},
  doi={10.1109/TNNLS.2018.2808470}}

@inproceedings{10.1145/3302504.3311807,
author = {Dutta, Souradeep and Chen, Xin and Sankaranarayanan, Sriram},
title = {Reachability analysis for neural feedback systems using regressive polynomial rule inference},
year = {2019},
isbn = {9781450362825},
publisher = {Association for Computing Machinery},
address = {New York, NY, USA},
url = {https://doi.org/10.1145/3302504.3311807},
doi = {10.1145/3302504.3311807},
booktitle = {Proceedings of the 22nd ACM International Conference on Hybrid Systems: Computation and Control},
pages = {157–168},
numpages = {12},
location = {Montreal, Quebec, Canada},
series = {HSCC '19}
}

@article{10.1007/s10817-018-09509-5,
author = {Dreossi, Tommaso and Donz\'{e}, Alexandre and Seshia, Sanjit A.},
title = {Compositional Falsification of Cyber-Physical Systems with Machine Learning Components},
year = {2019},
issue_date = {Dec 2019},
publisher = {Springer-Verlag},
address = {Berlin, Heidelberg},
volume = {63},
number = {4},
issn = {0168-7433},
url = {https://doi.org/10.1007/s10817-018-09509-5},
doi = {10.1007/s10817-018-09509-5},
journal = {J. Autom. Reason.},
month = dec,
pages = {1031–1053},
numpages = {23}
}

@inproceedings{10.1007/978-3-031-37703-7_19,
author = {Lopez, Diego Manzanas and Choi, Sung Woo and Tran, Hoang-Dung and Johnson, Taylor T.},
title = {NNV 2.0: The Neural Network Verification Tool},
year = {2023},
isbn = {978-3-031-37702-0},
publisher = {Springer-Verlag},
address = {Berlin, Heidelberg},
url = {https://doi.org/10.1007/978-3-031-37703-7\_19},
doi = {10.1007/978-3-031-37703-7\_19},
booktitle = {Computer Aided Verification: 35th International Conference, CAV 2023, Paris, France, July 17–22, 2023, Proceedings, Part II},
pages = {397–412},
numpages = {16},
location = {Paris, France}
}

@inproceedings{10.5555/2958031.2958095,
author = {Donz\'{e}, Alexandre and Ferr\`{e}re, Thomas and Maler, Oded},
title = {Efficient Robust Monitoring for STL},
year = {2013},
isbn = {9783642397981},
publisher = {Springer-Verlag},
address = {Berlin, Heidelberg},
booktitle = {Proceedings of the 25th International Conference on Computer Aided Verification - Volume 8044},
pages = {264–279},
numpages = {16},
location = {Saint Petersburg, Russia},
series = {CAV 2013}
}

@InProceedings{10.1007/978-3-030-25540-4_25,
author="Dreossi, Tommaso
and Fremont, Daniel J.
and Ghosh, Shromona
and Kim, Edward
and Ravanbakhsh, Hadi
and Vazquez-Chanlatte, Marcell
and Seshia, Sanjit A.",
editor="Dillig, Isil
and Tasiran, Serdar",
title="VerifAI: A Toolkit for the Formal Design and Analysis of Artificial Intelligence-Based Systems",
booktitle="Computer Aided Verification",
year="2019",
publisher="Springer International Publishing",
address="Cham",
pages="432--442",
isbn="978-3-030-25540-4"
}

@InProceedings{stl,
author="Maler, Oded
and Nickovic, Dejan",
editor="Lakhnech, Yassine
and Yovine, Sergio",
title="Monitoring Temporal Properties of Continuous Signals",
booktitle="Formal Techniques, Modelling and Analysis of Timed and Fault-Tolerant Systems",
year="2004",
publisher="Springer Berlin Heidelberg",
address="Berlin, Heidelberg",
pages="152--166",
isbn="978-3-540-30206-3"
}

@inproceedings{autolirpa,
author = {Xu, Kaidi and Shi, Zhouxing and Zhang, Huan and Wang, Yihan and Chang, Kai-Wei and Huang, Minlie and Kailkhura, Bhavya and Lin, Xue and Hsieh, Cho-Jui},
title = {Automatic perturbation analysis for scalable certified robustness and beyond},
year = {2020},
isbn = {9781713829546},
publisher = {Curran Associates Inc.},
address = {Red Hook, NY, USA},
booktitle = {Proceedings of the 34th International Conference on Neural Information Processing Systems},
articleno = {96},
numpages = {13},
location = {Vancouver, BC, Canada},
series = {NIPS '20}
}

@inproceedings{bernnn,
author = {Fatnassi, Wael and Khedr, Haitham and Yamamoto, Valen and Shoukry, Yasser},
title = {BERN-NN: Tight Bound Propagation For Neural Networks Using Bernstein Polynomial Interval Arithmetic},
year = {2023},
isbn = {9798400700330},
publisher = {Association for Computing Machinery},
address = {New York, NY, USA},
url = {https://doi.org/10.1145/3575870.3587126},
doi = {10.1145/3575870.3587126},
booktitle = {Proceedings of the 26th ACM International Conference on Hybrid Systems: Computation and Control},
articleno = {19},
numpages = {11},
location = {San Antonio, TX, USA},
series = {HSCC '23}
}

@article{reachnn,
author = {Huang, Chao and Fan, Jiameng and Li, Wenchao and Chen, Xin and Zhu, Qi},
title = {ReachNN: Reachability Analysis of Neural-Network Controlled Systems},
year = {2019},
issue_date = {October 2019},
publisher = {Association for Computing Machinery},
address = {New York, NY, USA},
volume = {18},
number = {5s},
issn = {1539-9087},
url = {https://doi.org/10.1145/3358228},
doi = {10.1145/3358228},
journal = {ACM Trans. Embed. Comput. Syst.},
month = oct,
articleno = {106},
numpages = {22}
}

@INPROCEEDINGS{kohei,
  author={Tsujio, Kohei and Al Faruque, Mohammad Abdullah and Shoukry, Yasser},
  booktitle={2024 ACM/IEEE 15th International Conference on Cyber-Physical Systems (ICCPS)}, 
  title={Rampo: A CEGAR-based Integration of Binary Code Analysis and System Falsification for Cyber-Kinetic Vulnerability Detection}, 
  year={2024},
  volume={},
  number={},
  pages={45-54},
  doi={10.1109/ICCPS61052.2024.00011}}

@inproceedings{z3,
author = {De Moura, Leonardo and Bj\o{}rner, Nikolaj},
title = {Z3: An Efficient SMT Solver},
year = {2008},
isbn = {3540787992},
publisher = {Springer-Verlag},
address = {Berlin, Heidelberg},
booktitle = {Proceedings of the Theory and Practice of Software, 14th International Conference on Tools and Algorithms for the Construction and Analysis of Systems},
pages = {337–340},
numpages = {4},
location = {Budapest, Hungary},
series = {TACAS'08/ETAPS'08}
}

@inproceedings{angr,
  title={{SoK: (State of) The Art of War: Offensive Techniques in Binary Analysis}},
  author={Shoshitaishvili, Yan and Wang, Ruoyu and Salls, Christopher and
          Stephens, Nick and Polino, Mario and Dutcher, Audrey and Grosen, John and
          Feng, Siji and Hauser, Christophe and Kruegel, Christopher and Vigna, Giovanni},
  booktitle={IEEE Symposium on Security and Privacy},
  year={2016}
}

@InProceedings{s-taliro,
author="Annpureddy, Yashwanth
and Liu, Che
and Fainekos, Georgios
and Sankaranarayanan, Sriram",
editor="Abdulla, Parosh Aziz
and Leino, K. Rustan M.",
title="S-TaLiRo: A Tool for Temporal Logic Falsification for Hybrid Systems",
booktitle="Tools and Algorithms for the Construction and Analysis of Systems",
year="2011",
publisher="Springer Berlin Heidelberg",
address="Berlin, Heidelberg",
pages="254--257",
isbn="978-3-642-19835-9"
}

@article{de2021symbolic,
  title={Symbolic execution formally explained},
  author={de Boer, Frank S and Bonsangue, Marcello},
  journal={Formal Aspects of Computing},
  pages={1--20},
  year={2021},
  publisher={Springer}
}

@inproceedings{maler2004monitoring,
  title={Monitoring temporal properties of continuous signals},
  author={Maler, Oded and Nickovic, Dejan},
  booktitle={International Symposium on Formal Techniques in Real-Time and Fault-Tolerant Systems},
  pages={152--166},
  year={2004},
  organization={Springer}
}

@inproceedings{10.1007/978-3-642-14295-6_17,
author = {Donz\'{e}, Alexandre},
title = {Breach, a toolbox for verification and parameter synthesis of hybrid systems},
year = {2010},
isbn = {364214294X},
publisher = {Springer-Verlag},
address = {Berlin, Heidelberg},
url = {https://doi.org/10.1007/978-3-642-14295-6\_17},
doi = {10.1007/978-3-642-14295-6\_17},
booktitle = {Proceedings of the 22nd International Conference on Computer Aided Verification},
pages = {167–170},
numpages = {4},
location = {Edinburgh, UK},
series = {CAV'10}
}

@inproceedings{stl2nn,
author = {Hashemi, Navid and Hoxha, Bardh and Yamaguchi, Tomoya and Prokhorov, Danil and Fainekos, Georgios and Deshmukh, Jyotirmoy},
title = {A Neurosymbolic Approach to the Verification of Temporal Logic Properties of Learning-enabled Control Systems},
year = {2023},
isbn = {9798400700361},
publisher = {Association for Computing Machinery},
address = {New York, NY, USA},
url = {https://doi.org/10.1145/3576841.3585928},
doi = {10.1145/3576841.3585928},
booktitle = {Proceedings of the ACM/IEEE 14th International Conference on Cyber-Physical Systems (with CPS-IoT Week 2023)},
pages = {98–109},
numpages = {12},
location = {San Antonio, TX, USA},
series = {ICCPS '23}
}

@inproceedings{verisig,
author = {Ivanov, Radoslav and Weimer, James and Alur, Rajeev and Pappas, George J. and Lee, Insup},
title = {Verisig: verifying safety properties of hybrid systems with neural network controllers},
year = {2019},
isbn = {9781450362825},
publisher = {Association for Computing Machinery},
address = {New York, NY, USA},
url = {https://doi.org/10.1145/3302504.3311806},
doi = {10.1145/3302504.3311806},
booktitle = {Proceedings of the 22nd ACM International Conference on Hybrid Systems: Computation and Control},
pages = {169–178},
numpages = {10},
location = {Montreal, Quebec, Canada},
series = {HSCC '19}
}

@InProceedings{reachnn_star,
author="Fan, Jiameng
and Huang, Chao
and Chen, Xin
and Li, Wenchao
and Zhu, Qi",
editor="Hung, Dang Van
and Sokolsky, Oleg",
title="ReachNN*: A Tool for Reachability Analysis of Neural-Network Controlled Systems",
booktitle="Automated Technology for Verification and Analysis",
year="2020",
publisher="Springer International Publishing",
address="Cham",
pages="537--542",
isbn="978-3-030-59152-6"
}

@inproceedings{DeepBern,
author = {Khedr, Haitham and Shoukry, Yasser},
title = {DeepBern-Nets: taming the complexity of certifying neural networks using bernstein polynomial activations and precise bound propagation},
year = {2024},
isbn = {978-1-57735-887-9},
publisher = {AAAI Press},
url = {https://doi.org/10.1609/aaai.v38i19.30117},
doi = {10.1609/aaai.v38i19.30117},
booktitle = {Proceedings of the Thirty-Eighth AAAI Conference on Artificial Intelligence and Thirty-Sixth Conference on Innovative Applications of Artificial Intelligence and Fourteenth Symposium on Educational Advances in Artificial Intelligence},
articleno = {2369},
numpages = {9},
series = {AAAI'24/IAAI'24/EAAI'24}
}

@inproceedings{10.5555/2032305.2032335,
author = {Frehse, Goran and Le Guernic, Colas and Donz\'{e}, Alexandre and Cotton, Scott and Ray, Rajarshi and Lebeltel, Olivier and Ripado, Rodolfo and Girard, Antoine and Dang, Thao and Maler, Oded},
title = {SpaceEx: scalable verification of hybrid systems},
year = {2011},
isbn = {9783642221095},
publisher = {Springer-Verlag},
address = {Berlin, Heidelberg},
booktitle = {Proceedings of the 23rd International Conference on Computer Aided Verification},
pages = {379–395},
numpages = {17},
location = {Snowbird, UT},
series = {CAV'11}
}

@inproceedings{flow*,
author = {Chen, Xin and \'{A}brah\'{a}m, Erika and Sankaranarayanan, Sriram},
title = {Flow*: An Analyzer for Non-linear Hybrid Systems},
year = {2013},
isbn = {9783642397981},
publisher = {Springer-Verlag},
address = {Berlin, Heidelberg},
booktitle = {Proceedings of the 25th International Conference on Computer Aided Verification - Volume 8044},
pages = {258–263},
numpages = {6},
location = {Saint Petersburg, Russia},
series = {CAV 2013}
}

@InProceedings{verisig2,
author="Ivanov, Radoslav
and Carpenter, Taylor
and Weimer, James
and Alur, Rajeev
and Pappas, George
and Lee, Insup",
editor="Silva, Alexandra
and Leino, K. Rustan M.",
title="Verisig 2.0: Verification of Neural Network Controllers Using Taylor Model Preconditioning",
booktitle="Computer Aided Verification",
year="2021",
publisher="Springer International Publishing",
address="Cham",
pages="249--262",
isbn="978-3-030-81685-8"
}

@InProceedings{nnv,
author="Tran, Hoang-Dung
and Yang, Xiaodong
and Manzanas Lopez, Diego
and Musau, Patrick
and Nguyen, Luan Viet
and Xiang, Weiming
and Bak, Stanley
and Johnson, Taylor T.",
editor="Lahiri, Shuvendu K.
and Wang, Chao",
title="NNV: The Neural Network Verification Tool for Deep Neural Networks and Learning-Enabled Cyber-Physical Systems",
booktitle="Computer Aided Verification",
year="2020",
publisher="Springer International Publishing",
address="Cham",
pages="3--17",
isbn="978-3-030-53288-8"
}

\end{document}